\documentclass[preprint,preprintnumbers]{revtex4}

\usepackage{graphicx}

\usepackage{amssymb}

\usepackage{mathrsfs}

\begin{document}

\title{Propagation  of Phonons  in  a Curved Space Induced  by  Strain Fields Instantons} 

\author{D. Schmeltzer}

\affiliation{Physics Department, City College of the City University of New York,  
New York, New York 10031, USA}

\begin{abstract}
We show that for  a \textbf{multiple-connected}   space   the low energy    strain fields excitations    are given  by instantons.  Dirac fermions with  a chiral  mass and a pairing field propagates effectively in a multiple conected space. When   the  elastic  strain field  response is probed one finds that it is  given by  the \textbf{Pointriagin} characteristic.  As a result   the space  time metric is modified.  Applying an external stress  field  we  observe  that the  phonon path   bends in the transverse direction to the initial direction.

\end{abstract}

\maketitle

\vspace{0.2 in}

\textbf{I- Introduction}

\vspace{0.2 in}

 Gauge theories and their applications  to gravity and  elasticity  are  important  tools  in  modern  physics \cite{Kleinert,Kleinertnew}.
In gauge theories the gauge potential is  fundamental and  determines  the field strength.
The same approach is used in gravity where the \textbf{vierbein} (the transformation matrix of the coordinates  )  determines the  \textbf{spin-connection} and the \textbf{curvature} \cite{Nieh,Kleinert}. The  similarity between  the theory of gravity and elasticity has been pointed out by 
Kleinert \cite{Kleinert}.  The    classical theory of  elasticity \cite{Lifshitz} is based  on the formalism  of coordinates transformations between  the perfect and the deformed  crystal.  As a result the space derrivatives are replaced by covariant derivatives  which introduce  the concept of \textbf{spin-connection}.  The    disclinations density  have  been  described by   the spatial part of  the Einstein tensor and the dislocation density  are  given by the  torsion tensor \cite{Katanev,Kleinert}.
The \textbf{spin-connection} 
might play a the same  fundamental role in elasticity as the  gauge fields in gauge  theories. This   suggests  the possibility  of \textbf{instantons }  in elasticity.  
This  possibility follows from    Volterra's work \cite{Volterra}, which used the  homotopy method  to describe   disclinations and dislocations .  The Saint Venat  compatibility  conditions   imply that the torsion and curvature must vanish \cite{Volterra,Kleinert}.
 Volterra concluded    that the equilibrium state of a simply connected elastic body is a state with zero strain, while the equilibrium state of a \textbf{multiple-connected} one could have non-vanishing kinetic energy with non-vanishing strain \cite{Delphenich}.
Such a state corresponds to the  existence of quantum instantons, similar to the metastable states for two dimensional isotropic ferromagnet and   four dimensional Yang-Mills  fields \cite{Yang-Mills}.
 In order  to realize the physics of instantons in elasticity,  one needs a  \textbf{multiple-connected}  space.  
For example when  the  Dirac Hamiltonian contains a chiral  mass  $\gamma_{5}m$, the model will have zero modes and therefore is  topological non-trivial. If instead of  probing the Dirac Hamiltonian   with an external   electromagnetic  field one  probes the system   with a  strain field   the  response will  generates   the   \textbf{Pointriagin} characteristic  \cite{Nakahara}. This response is similar to the one in   electromagnetism   where  the Dirac Hamiltonian with a chiral mass  $\gamma_{5}m$ modifies the Maxwell's equation  by the  term 
  $\delta H\propto\theta \vec{E}\cdot\vec{B}$ ( $\theta$ represents the axion field \cite{Wilczek}).  Such a term 
 gives rise  to  the  Faraday and Kerr rotation \cite{Karch,Tse,Obukhov,davidM}.  

Inspired by the analogy between the gauge theory and the theory of  elasticity \cite{Kleinert,Nieh} we expect an  elastic analog  to   the magnetoelectic effect.  In elasticity the   phonon replaces the  photon therefore,  an analog to the Farady effect  is expected. This suggests that a medium  which is described by an elastic  topological invariant  will   affect the  phonon  propagation and   polarization.     In elasticity a local crystal  deformation induces a \textbf{spin-connection}  which has  curvature  (disclinations) and   an elastic \textbf{ Pointriagin} invariant.    For  a Dirac Hamiltonian with a chiral mass term  $\gamma_{5}m$  and a pairing field  one obtains     a  topological superconductor.  For such a  case we can use  the elastic response  to  identify   the  topological superconductor. 
The topological invariants in elasticity  are   given by their analog in gravity, the \textbf{Pointriagin} and the \textbf{Euler} characteristic (see eqs.$(3,4)$ in ref.\cite{Yang-Mills} ).
For the Yang-Mills gauge theory \cite{YGGR}  the \textbf{Pointriagin} electromagnetic  characteristics  \cite{YGGR} are known to give rise to the Yang-Mills   \textbf{instantons}  excitations.

In the  continuum theory of elasticity the equilibrium position of the atoms in a   solid  \cite{Lifshitz}  is given  by the  coordinate $\vec{r}$ . A  small  crystal   perturbation  is  described  by  the elastic  field   $u^{a}(\vec{r},t) $ ($a=1,2,3$ represents the Cartesian coordinates). Due to the periodicity of the elastic    field,  the deviation from equilibrium $u^{a}(\vec{r},t)$ and  $u^{a}(\vec{r}+\vec{a},t)$ describes  the same physical excitation  ($\vec{a}$ represents the crystal Bravais  lattice) \cite{Kleinert}. 
Due to   the presence of the electron lattice  coupling  
one  expects that the lattice strain field $u^{a}(\vec{r},t) $  defined in the absence of  superconductivity  will be   modified once  superconductivity   is  switched  on .  As a result the coordinate $\vec{r}$  must be replaced  by  a new  lattice coordinate,   $\vec{r}\rightarrow \vec{r}+\vec{e}(\vec{r},t)$.  When the lattice coordinates are   modified     the  displacement  field   $u^{a}(\vec{r},t) $  is replaced  by  the   field   $u^{a}(\vec{e}(\vec{r},t)) $ which induces  new    strain fields. 
 Following  ref. \cite{Nieh,Kleinert}   the derivatives   with respect  the \textbf{Lorentz}  coordinates  $\mu=t,x,y,z $   \cite{Kleinert}  are replaced by the covariant derivative  $\nabla_{\mu}$ with the \textbf{spin-connection} $A^{a,b}= A^{a,b}_{\mu}dx^{\mu}$. The  covariant derivative of the  elastic field  $u^{a}(\vec{r},t) $  and  the electronic  spinor $\Psi$  are  given by : $\nabla_{\mu}u^{a}(\vec{r},t) =\partial_{\mu}u^{a}(\vec{r},t)+A^{a,b}_{\mu}u^{b}(\vec{r},t)$ and 
 $\nabla_{\mu}\Psi(\vec{r},t)=\partial_{\mu}\Psi(\vec{r},t)+\frac{1}{4}A_{\mu}^{a,b}[\gamma^{a},\gamma^{b}]\Psi$  where   $a=0,1,2,3$  and  $ \gamma^{a}$ are the Dirac  matrices .

The integration of the Dirac fermion in the presence of  the term   $\gamma_{5}m $  induces   the \textbf{Pointriagin} characteristic \cite{Richter,Egouchi,Freund}   $\int d^4x\sqrt{-g}[\frac{1}{768 \pi^2}\epsilon^{\mu,\nu,\alpha,\beta}F^{a}_{\alpha,\beta,b}F^{a}_{\mu,\nu,b}]$ given in terms of the metric tensor $g_{\mu,\nu}$  and \textbf{ elastic field strength}  $F_{\mu,\nu,b}^{a}=\partial_{\mu}A_{\nu}^{a,b}-\partial_{\nu}A_{\mu}^{a,b}+[A_{\mu},A_{\mu}]^{a}_{b}
$.
From  the work of  \cite{Katanev} we learn that the elastic deformations  which describe the  topological defects  is given by a term which is  similar to the  Hilbert-Einstein  action (see eq.$(30)$  in ref. \cite{Katanev}.   The variation of the elastic action  is satisfied for elastic field strengths  which satisfy    the     duality or  anti-duality  conditions $\tilde{A}^{a,b}=\pm A^{a,b} $ From  the analogy with the Yang-Mills gauge fields we learn that the  instanton field   obeys  the     duality or  anti-duality  conditions   $\tilde  { F}_{\mu,\nu}^{a,b}=\pm  F_{\mu,\nu}^{a,b} $ , $\tilde  { F}_{\mu,\nu,b}^{a,b} \equiv \frac{1}{2}\sum_{c=0,d=0}^{c=3,d=3}\epsilon_{abcd} F_{\mu,\nu}^{c,d}$.  Each solution is satisfied for an arbitrary integration  constant $\Lambda$. For  each  $\Lambda$ the  \textbf{Pointriagin} action   gives the   instanton weight, $e^{-\frac{\Lambda^{2}}{const}}$ in agreement  \cite{Egouchi,Freund}.( As a result 
 the metric    satisfy    at long distances the conditions   $g_{\mu,\nu}\rightarrow 0$  with      the field strength    $F^{\alpha}_{\beta,\mu,\nu}$  and the  Ricci curvature   $F_{\mu,\nu} $   obey:$ F^{\alpha}_{\beta,\mu,\nu}\rightarrow 0$ ,     $F_{\mu,\nu}\rightarrow 0$ . Only  the scalar curvature $g^{\mu,\mu}F_{\mu,\mu}$ is finite \cite{Egouchi}.) 

The purpose of this paper is to investigate the   propagation of the phonon field in a space controlled by  the  metric tensor $g_{\mu,\nu}$  induced by the instantons.
Using the  Euclidean  version  \textbf{sound} of the  Lorentz transformation we describe the unperturbed crystal  which  as  $S^3$ Euclidean space.   The effective action of the crystal  is  given  by the elasticity  action  \cite{Katanev,Kleinert}  which is similar to the gravitation Einstein action.   A   topological superconductor   for a Dirac Hamiltonian with pairing interaction which is controlled by the crystal  deformation, the presence of a term of the form $\gamma_{5}$  coupled to the order parameter of the Superconductor  Superconductor   gives rise to the \textbf{Pointriagin}  action.   controlled by the saddle point of the Superconducting order parameter.  (The elastic  deformed crystal affects the  electronic action   by generating   an elastic term  which given  by  the   Pointriagin   characteristic in elasticity  \cite{ Richter,YGGR}).
The possible elastic  deformation of the  Euclidean $ S^3$ sphere which obey the elastic version of the  Hilbert-Einstein  action  \cite{Katanev}  satisfies  the duality or   anti-duality relations.   Each deformation of the crystal   corresponds to  a  metric solution  with a different  integration factors  $\Lambda$.  For each  $\Lambda$ we have an instanton contribution  to the action $e^{-\frac{\Lambda^{2}}{const}}$.  The maximum  contribution to the action comes from the unperturbed  crystal  with  $\Lambda=0$.  The propagation  of the phonon is affected by the metric tensor.  For each value $\Lambda$ we have a different metric which will affect   the phonon propagation.  Therefore the phonon propagation is obtained after  we  add all the contributions.  In order to enhance  the statistical weight  for values of   $\Lambda\neq0$ we can  bend the crystal   in the superconducting phase. 
  As a concrete example  we compute the metric tensor for a fixed value of  $\Lambda$ . Using the computed metric tensor  we compute numerically  the propagation of  the phonon at long distances and   find a  bended  path .
 
The plan of this paper is as follows. In Sec. $II$  we introduce the \textbf{Pointriagin} index in gravity and propose that such an index emerges also in elasticity under certain conditions.
Sec.$III$ presents the emergence of the \textbf{Pointriagin} index  in elasticity. 
In addition the elastic energy of the crystal gives rise to a term similar to the Einstein curvature or  density.  In Sec.$IV$ we show that the Green's function of the lattice is a function of the metric. Therefore  the propagation of the elastic waves is obtained after performing a statistical average over the metric. Sec.$V$ is devoted to computation of the metrics.  The spin connection needs to satisfy self duality conditions. As a result one find that at long distance a metric characterized by the  integration  factor  $\Lambda$. In Sec.$VI$ we  consider a simplified version of the problem and show that for a single value of $\Lambda\neq 0$  we obtain at long distances a bended path for  the  propagating phonon.

\vspace{0.2 in}

\textbf{II-The  Pointriagin index in electromagnetism and gravity}

\vspace{0.2 in}

The \textbf{Pointriagin} index \textbf{$\nu$}  measures the difference between the vanishing eigenvalues of positive  and negative chirality of the Euclidean Dirac operator  $\gamma^{\mu}D_{\mu}[\vec{A}]$ (in the presence of the electromagnetic field $\vec{A}$) with eigenvalues $\gamma_{5}=\pm 1$ \cite {Nakahara}. 
\begin{equation}
I_{P}=\nu\equiv n_{+}- n_{-}=\frac{1}{16\pi^2}\int d^4xTr[\tilde{F}_{\mu,\nu}  F^{\mu,\nu}]=-\frac{1}{14\pi^2}\int d^4x [\vec{E}\cdot\vec{B}]
\label{mag}
\end{equation}
$\tilde{F}_{\mu,\nu}$ is the dual strength field  ($ \vec{B}  $)  of  $ F^{\mu,\nu}$  (the  field $ \vec{E}$).
The  \textbf{Pointriagin} index has  been  defined   in gravity   \cite{Egouchi,Richter} and also used for Topological Superconductors \cite{Ludwig,Wang} where  $A^{a,b}_{\mu}$ is the   \textbf{spin- connection} .
\begin{equation}
I_{p,el.}=\nu\equiv n_{+}- n_{-}=
-\int d^4x\sqrt{-g}[\frac{1}{768 \pi^2}\epsilon^{\mu,\nu,\alpha,\beta}F^{a}_{\alpha,\beta,b}F^{a}_{\mu,\nu,b}];\hspace{0.1 in}g\equiv det[g_{\mu,\nu}]
\label{index}
\end{equation}
\label{mag}
  We will show in the next section  that the effect of  strain  fields in     elasticity  coupled to a pairing field generates   a Pointriagin  term   as encountered in gravity.  \cite{Katanev}. In a crystal   the presence of an external stress or defects such as dislocations or disclinations induce a  \textbf{strain} field  \cite{Katanev}.  The  strain field  gives  rise to a \textbf{spin-connection}   $A^{a,b}_{\mu}$  with the    \textbf{elastic field strength}  $ F^{a}_{\alpha,\beta,b}$ \cite{Katanev,Kleinert}.

\vspace{0.2 in}

 \textbf{III- Construction of the effective action   for a Dirac  Hamiltonian with a pairing interaction     coupled to   elasticity}

\vspace{0.2 in}
 
The  \textbf{Pointriagin} index in elasticity is obtained   for  a   Dirac Hamiltonian  with a pairing interaction coupled to the  crystal  strain  field  in the superconducting phase. 
The    Dirac Hamiltonian \cite{Chamon}  with a paring interaction
 $\varphi$ and $\varphi^{*}$ is a function crystal field  $\vec{u}(\vec{r},t)$.  
\begin{eqnarray}
&&H^{s.c}=\int_{-\infty}^{\infty}\, dt\int d^3r\Big[[\psi^{\dagger}_{\uparrow}(\vec{r},t),\psi^{\dagger}_{\downarrow}(\vec{r},t)]^T(\sigma^{1}(-i)\partial_{1}+\sigma^{2}(-i)\partial_{2})[\psi_{\uparrow}(\vec{r},t),\psi_{\downarrow}(\vec{r},t)]\nonumber\\&&+\varphi(\vec{r},t) \psi^{\dagger}_{\uparrow}(\vec{r},t)\psi^{\dagger}_{\downarrow}(\vec{r},t)+\varphi^{*}(\vec{r},t)\psi_{\downarrow}(\vec{r},t)\psi_{\uparrow}(\vec{r},t)\Big] -\int_{-\infty}^{\infty} dt\int d^3r \varphi^{*}(\vec{r},t)\varphi(\vec{r},t)\nonumber\\&&
\nonumber\\&&
\end{eqnarray}
We introduce a four component spinor
\begin{equation}
\Psi\equiv[\psi_{\uparrow},\psi_{\downarrow},\psi^{*}_{\downarrow},-\psi^{*}_{\uparrow}]^{T} 
\label{spinor}
\end{equation}
The spinor $\Psi$ obeys :
$\Psi_{C}=C \Psi^{*}=\Psi$; \hspace{0.1 in} $C=i\sigma^2\times\tau_{2}$
 where   $ \sigma^2$ is the   spin   Pauli matrix   and $C=i\sigma^2\times\tau_{2}$  is  charge conjugation matrix. 
Using the constrained spinor $\Psi=C \Psi^{*}$ we obtain the  Dirac Hamiltonian  with pairing interaction \cite{Chamon} .
\begin{eqnarray}
&&S[\bar{\Psi},\Psi,\varphi, \vec{ u}(\vec{r},t)]=\int_{-\infty}^{\infty} dt\int d^3r
\Big[\bar{\Psi}(\vec{r},t)[i\gamma^{0}\partial_{t}+i\gamma^{1}\partial_{1}+i\gamma^{2}\partial_{2} +i\gamma^{3}\partial_{3}+\varphi_{r}+ i\gamma^{5}\varphi_{I} ]\Psi(\vec{r},t)\Big]\nonumber\\&&-\frac{1}{2}\int_{-\infty}^{\infty} dt\int d^3r (\varphi^2_{r}(\vec{r},t)+\varphi^2_{I}(\vec{r},t))+\int_{-\infty}^{\infty} dt\int d^3r
\frac{1}{2}[\partial_{t}(u^{a}(\vec{r},t))^2-\lambda_{c,a,d,b}\partial_{c}(u^{a}(\vec{r},t))\partial_{d}(u^{b}(\vec{r},t))]\nonumber\\&&
\end{eqnarray}
 $\varphi_{r}$ and  $\varphi_{I}$ are the real and imaginary  pairing field.
We have included in eq.$(5)$ ( the last term) the   crystal action in the $harmonic$ approximation  \cite{Lifshitz}  with  the Lame coefficients
$\lambda_{\mu,a,\nu,b}$   \cite{Lifshitz},
$\lambda_{c,a,d,b}=(K-\frac{2}{3}\mu)\delta_{c,a}\delta_{d,b}+G(\delta_{c,d}\delta_{a,b}+\delta_{c,b}\delta_{a,d})$. $K$ is the compressibility and $\mu$ is the  shear  modulus of the crystal.
When the  pairing field  vanish  $\varphi=0$,  $\varphi^{*}=0$  the Green's function  for  the lattice  field    $u^{a}(\vec{r},t)$ is given by:
$G^{a,b}(\vec{r},\vec{r'};t-t')=-i\langle 0|T\Big[u^{a}(\vec{r},t)  u^{b}(\vec{r'},t')\Big]|0\rangle $.
Where $|0\rangle$ is the unperturbed crystal and $T $ is the time order operator  \cite{Kleinert}. 
When  $\varphi=\varphi^{*}=0 $ the crystal is described by  the  coordinate $ t$ and $\vec{r}$  which 
satisfy  the  sound \textbf{ Lorentz}  transformation,
$ds^2=-c^{2}dt^2+dx^2+dy^2+dz^2$,  $c=1$ is the speed of sound (transversal or longitudinal).
 In the superconducting phase  the pairing fields  $\varphi\equiv\varphi(\vec{ u}(\vec{r},t))$ , $\varphi^{*}(\vec{ u}(\vec{r},t))$  generates  the  mass term   $\gamma_{5}m$   ($m\equiv<\varphi_{r}>=<\varphi+\varphi^{*}>$) responsible for the Pointriagin term. These terms are function of the crystal strain field $u^{a}(\vec{r},t)$  and  the  external stress .  In particular when external defects such as \textbf{dislocations} or \textbf{disclination}  \cite{Kleinert,Katanev,davidNJP} are introduced  in the superconducting phase   the  new coordinates will be  given by $\vec{W}(\vec{r},t)$.
The deformed  coordinates induced by  the elastic deformations       obeys    the sound \textbf{ Lorentz}  transformation,  with the new coordinates $  [t'(\vec{r},t),\vec{W}(\vec{r},t)]\equiv[t'(\vec{r},t),\vec{r}+\vec{e}(\vec{r},t)]$.  This  transformation  replaces   crystal displacement field    $\vec{u}(\vec{r},t)$  by $\vec{u}(\vec{W}(\vec{r},t))$.

Following   \textbf{Cartan} formalism \cite{Nieh} we  incorporate the coordinate transformations with the help of the \textbf{vierbein}  matrix $e^{a}_{\mu}(\vec{r},t)\equiv \partial_{\mu} W^{a}(\vec{r},t)$    and the metric tensor  $g_{\mu,\nu}$.
\begin{eqnarray}
 &&e^{a}_{\mu}(\vec{r},t)\equiv \partial_{\mu} W^{a}(\vec{r},t);a=0, 1,2,3;\mu=t,x,y,z\nonumber\\&& 
g_{\mu,\nu}=\sum_{a=0}^{3}e^{a}_{\mu}e^{a}_{\nu}\nonumber\\&&
\sum_{\mu}e^{a}_{\mu}e^{\mu}_{b}=\eta_{a,b}
\nonumber\\&&
\end{eqnarray}  
$ \eta_{a,b}=(-1,1,1,1)$,  for the Euclidean case  $\eta_{a,b}$ is replaced by $\delta_{a,b}$
Using the   matrix   $e^{a}_{\mu}$  we describe  the transformation  from   $local$  orthonormal coordinates $ a= 0,1,2,3$  to  the  fixed   Cartesian frame   (given by the  Greek letters)  $\mu=t,x,y,z$.
Following  ref. \cite{Nieh,davidNJP} we  obtain the elastic \textbf{spin-connection} $A^{a,b}_{\mu}$ 
\begin{eqnarray}
&&A_{\mu}^{a,b}=\frac{1}{2}[e^{\nu,a}(\partial_{\mu}.e^{b}_{\nu}-\partial_{\nu}e^{b}_{\mu})-e^{\nu,b}(\partial_{\mu}e^{a}_{\nu}-\partial_{\nu}e^{a}_{\mu})-e^{\rho,a}e^{\sigma,a}(\partial_{\rho}e_{\sigma,a}-\partial_{\sigma}e_{\rho,c})e^{c}_{\nu}]\nonumber\\&&
e^{a,\nu}\equiv e^{a}_{\sigma}g^{\sigma,\nu}\nonumber\\&&
\end{eqnarray}
 The covariant derivative of the  spinor $ \Psi$ with the four components  $\Psi^{a}(\vec{r},t)$, $a=0,2,3,4$ is given by:
\begin{eqnarray}
&&\nabla_{\mu}\Psi(\vec{r},t)=\partial_{\mu}\Psi(\vec{r},t)+\frac{1}{4}A_{\mu}^{a,b}[\gamma^{a},\gamma^{b}]\Psi
\nonumber\\&&
\end{eqnarray}
where $\gamma^{0}=I\times \tau_{3}$ and   $\gamma^{a} =\sigma^{a}\times i\tau_{2}$ for $a=1,2,3$.  $\tau $ the denotes the Pauli matrix in the $orbital$ or $chiral $ space and $I$, $\sigma^{a}$ represents the Pauli matrix in the $ SU(2)$ spin  space .
Due to the elastic strain field the  derivative of the elastic field $\vec{u}(\vec{r},t)$  is  replaced by the  covariant derivative  $\nabla_{\mu}$  given in terms  of the the same  \textbf{spin-connection} A$^{a,b}_{\nu}(\vec{r},t)$   which  determine  the covariant derivative of the fermion spinor.
\begin{equation}
\nabla_{\mu}u^{a}(\vec{r},t) =\partial_{\mu}u^{a}(\vec{r},t)+A^{a,b}_{\mu}u^{b}(\vec{r},t)
\label{covariant}
\end{equation}
As a result 
the \textbf{ strength field}  $F_{\mu,\nu,b}^{a} $   given in terms of the  \textbf{spin-connection} $A_{\mu}^{a,b} $.
\begin{equation}
F_{\mu,\nu,b}^{a}=\partial_{\mu}A_{\nu}^{a,b}-\partial_{\nu}A_{\nu}^{a,b}+[A_{\mu},A_{\mu}]^{a}_{b}
\label{curvature}
\end{equation}
The  curvature tensor $R_{\mu,\nu}^{\sigma,\tau}$,  the  Ricci tensor $R_{\mu,\nu}$ and the scalar curvature $R$ are given with the help of  the \textbf{spin-connection} $A_{\mu}^{a,b} $ and the \textbf{vierbein} matrix $e^{a}_{\mu}(\vec{r},t)$:
\begin{eqnarray}
&&R_{\mu,\nu}^{\lambda,\sigma}=e^{\lambda}_{a}e^{\sigma}_{b}
F_{\mu,\nu}^{a,b}\nonumber\\&&
F_{\mu,\nu,b}^{a}=e^{a}_{\sigma}e^{\tau}_{b}R_{\mu,\nu}^{\sigma,\tau}\nonumber\\&&
R_{\mu,\nu}= R_{\mu,\lambda,\nu}^{\lambda} =    e^{\lambda}_{a}e^{b}_{\lambda} F_{\mu,b,\nu}^{a}= \eta^{b}_{a} F_{\mu,b,\nu}^{a} =\epsilon_{\mu \alpha \beta}\epsilon_{\nu\rho \sigma}\partial_{\alpha}\partial_{\beta} \partial_{\rho}W_{\sigma}(\vec{r},t)\nonumber\\&&
R=g^{\nu,\mu}R_{\nu,\mu}= \eta^{b}_{a}  F_{\mu,b,\nu}^{a}\nonumber\\&&
\end{eqnarray}
 The  the \textbf{vierbein} will determine  the   \textbf{spin-connection}   the \textbf{field strength},\textbf{curvature}, \textbf{Ricci} tensor ,  \textbf{scalar curvature} and  the behavior of the  fields $ \Psi(\vec{r},t)$  and  $\vec {u}(\vec{r},t)$. 

and they affect the spinor field $ \Psi(\vec{r},t)$  and the  elastic  field  $\vec {u}(\vec{r},t)$. 

 The action  for the electrons and  lattice in the harmonic representation  $S[\bar{\Psi},\Psi,\varphi,\vec{u}]$ will be replaced by the \textbf{spin-connection} and \textbf{vierbein}:
\begin{equation}
S[\bar{\Psi},\Psi,\varphi,\vec{u}]\rightarrow \hat{S}[\bar{\Psi},\Psi,\varphi;e^{a}_{\mu},A^{a,b}_{\mu},\vec{u}]
\label{action}
\end{equation}
As a result the partition function will be given by :
\begin{equation}
Z=\int\,D \bar{\Psi} D\Psi D\varphi Du D e^{a}_{\mu} DA^{a,b}_{\mu}e^{i  \hat{S}[\bar{\Psi},\Psi,\varphi;e^{a}_{\mu},A^{a,b}_{\nu},\vec{u}]}
\label{partitionion}
\end{equation}
The integration  measure  $D e^{a}_{\mu} DA^{a,b}_{\mu}$  described all the possible  elastic deformation which can be generated  by the configuration $\vec{W}(\vec{r},t)$. The integration measure $Du$ describes the small deviation in the harmonic   approximation for different crystal configurations $\vec{W}(\vec{r},t)$.
The  explicit form of the action for  the deformed crystal  will be :
\begin{eqnarray}
&&\hat{S}[\bar{\Psi},\Psi,\varphi;e^{a}_{\mu},A^{a,b}_{\nu},\vec{u}]=\int_{-\infty}^{\infty} dt\int -d^3r\sqrt{det[g_{\mu,\nu}]}\Big[\bar{\Psi}(\vec{r},t)[i\gamma^{0} e^{\mu}_{0}(\partial_{\mu}+\frac{1}{8}A_{\mu}^{a,b}[\gamma^{a},\gamma^{b}])\nonumber\\&&+i\gamma^{1} e^{\mu}_{1}(\partial_{\mu}+\frac{1}{8}A_{\mu}^{a,b}+i\gamma^{2} e^{\mu}_{2}(\partial_{2} +\frac{1}{8}A_{\mu}^{a,b}[\gamma^{a},\gamma^{b}]) +i\gamma^{3} e^{\mu}_{3}(\partial_{\mu}+\frac{1}{8}A_{\mu}^{a,b}[\gamma^{a},\gamma^{b}])+\varphi_{r}(\vec{r},t)\nonumber\\&&+ i\gamma^{5}\varphi_{I}(\vec{r},t)] \Psi(\vec{r},t)\Big]-\frac{1}{2}\int_{-\infty}^{\infty} dt\int d^3r \sqrt{-det[g_{\mu,\nu}]} (\varphi_{r}^2(\vec{r},t)+\varphi_{I}^2(\vec{r},t) ) \nonumber\\&&
+\int_{-\infty}^{\infty} dt\int d^3r \sqrt{-det[g_{\mu,\nu}]}
\frac{1}{2}\Big[e^{\mu}_{0}\nabla_{\mu}(u^{a}(\vec{r},t))^2-\lambda_{c,a,d,b}e^{\mu}_{c}\nabla_{\mu}(u^{a}(\vec{r},t))e^{\nu}_{d}\nabla_{\nu}(u^{b}(\vec{r},t))\Big]\nonumber\\&&
\end{eqnarray}
We perform the chiral transformation and observe that the integration  measure is not  invariant \cite{Fujikawa}.
We integrate the Dirac Fermions action   in the presence of the chiral pairing field $i\gamma^{5}\varphi_{I}(\vec{r},t)$ and compute the    sadle point   with respect  $\varphi_{r}$  \cite{davidM}  and   find in agreement with  \cite{Egouchi,Richter}.
\begin{eqnarray}
&&\hat{S}_{eff.}[e^{a}_{\mu},A^{a,b}_{\nu},\vec{u}]= \int_{-\infty}^{\infty}\sqrt{-det[g_{\mu,\nu}]}\, dt \int_{-\infty}^{\infty}\, d^3r \Big[(\frac{\theta}{1536\pi^2}) \epsilon^{\mu,\nu,\alpha,\beta}F^{a}_{\alpha,\beta,b}F^{a}_{\mu,\nu,b}-\mathbf{E}(g,|\varphi|)\Big]\nonumber\\&&
+\int_{-\infty}^{\infty}\, dt\int_{-\infty}^{\infty}\, d^3r \sqrt{-det[g_{\mu,\nu}]}
\frac{1}{2}\Big[e^{\mu}_{0}\nabla_{\mu}(u^{a}(\vec{r},t))^2-\lambda_{c,a,d,b}e^{\mu}_{c}e^{\nu}_{d}\nabla_{\mu}(u^{a}(\vec{r},t))\nabla_{\nu}(u^{b}(\vec{r},t))\Big]\nonumber\\&&
 \theta=\frac{1}{2}tan^{-1}(\frac{\varphi_{I}}{\varphi_{r}})\nonumber\\&&
\end{eqnarray}
 The value of the  $\theta$ term is determined by the saddle point value for  the pairing field $\varphi$   as  in  ref. \cite{davidM}.
The the first term of the  action $\hat{S}_{eff.}$  is proportional to the   \textbf{Pointriagin} index $I_{p.el.}$ in eq. $(2)$ .   This term gives rise  to  $instantons$ excitations  similar to the one  obtained for     the Yang Mills  action  \cite{Egouchi,Yang-Mills,YGGR}.
The  second  term in equation  $(16)$ represents the  change in the  electronic energy due to the   the superconductor $\mathbf{E}(g,|\varphi|)\equiv\mathbf{E}(g)-c_{0}\varphi^2 >0$:
$\mathbf{E}(g)$ represents the energy of the free  Dirac Hamiltonian in the non-superconducting phase . For the superconducting case  the electronic energy is lowered   by the   pairing energy  $- c_{0}\varphi^2$. 
We observe that  the spin connection which determines the Pointriagin action  (the first term  in equation $(16)$   affects also the  equation of  motion for  the elastic  field $u^{a}(\vec{r},t)$.
The  variation of the elastic action      with respect   the elastic strain field $u^{a}(\vec{r},t)$,
$\frac{\delta S_{el.}}{\delta u^{a}(\vec{r},t)}=0$ gives the equation of motion for the sound waves in the deformed space.
It is possible to decompose $u^{a}(\vec{r},t)$ into the transversal field  $u^{a}_{T}(\vec{r},t)$ and the longitudinal one  $u^{a}_{L}(\vec{r},t)$:
\begin{equation}
 u^{a}(\vec{r},t)=u^{a}_{T}(\vec{r},t)+u^{a}_{L}(\vec{r},t)
\label{vec}
\end{equation}
 The transversal and   longitudinal fields  propagate with the sound velocities $c_{T}=\sqrt{\frac{\mu}{\rho}}$ (transversal) and 
$c_{L}=\sqrt{\frac{\lambda+2\mu}{\rho}}$ (longitudinal) .
 The effective action for the metric $g_{\mu,\nu}$ is obtained  as a result the  integration of the  field  $u^{a}(\vec{r},t)$. For simplicity we will approximate the dynamics of the crystal by   two scalar fields $ u_{s}(\vec{r},t)$   where $s=T,L$  this  will allow to replace the action by a scalar action  :
\begin{eqnarray}
&&\hat{S}_{eff.}[e^{a}_{\mu},A^{a,b}_{\nu},\vec{u}]= \int_{-\infty}^{\infty}\sqrt{-det[g_{\mu,\nu}]}\, dt \int_{-\infty}^{\infty}\, d^3r \Big[(\frac{\theta}{1536\pi^2}) \epsilon^{\mu,\nu,\alpha,\beta}F^{a}_{\alpha,\beta,b}F^{a}_{\mu,\nu,b}-\mathbf{E}(g,|\varphi|)\Big]\nonumber\\&&
+\int_{-\infty}^{\infty}\, dt\int_{-\infty}^{\infty}\, d^3r \sqrt{-det[g_{\mu,\nu}]}
\frac{1}{2}\Big  [ g^{\mu,\nu}\partial_{\mu} u_{s}(\vec{r},t)     \partial_{\nu} u_{s}(\vec{r},t)+ J_{s}(\vec{r},t)u_{s}(\vec{r},t)    \Big]\nonumber\\&&
 \theta=\frac{1}{2}tan^{-1}(\frac{\varphi_{I}}{\varphi_{r}})\nonumber\\&&
\end{eqnarray}
  $ J_{s}(\vec{r},t)$ is  the external current used to compute the Green's function.  Using the scalar approximation   we  perform the integration  over the  field $u_{s}(\vec{r},t)$  and find  the effective action.
 The  elasticity contribution to the effective action  for a  scalar field  was given in ref. \cite{Christensen} and more recently in elasticity \cite{Katanev}. There are two contributions ; $\kappa$ a  constant shift in the ground state energy and  a term  $B(\mu)R$  with  $B(\mu)$ being proportional to the shear modulus $\mu$ of the crystal and $R$ the scalar curvature. 
\begin{eqnarray}
&&S_{eff.}[e^{a}_{\mu},A^{a,b}_{\mu};J_{s}]\equiv S_{eff.}[e^{a}_{\mu},A^{a,b}_{\mu}]  +\int_{-\infty}^{\infty}\,dt  \int d^3r  \int_{-\infty}^{\infty}\,dt'  \int d^3r'\Big[ J_{s}(\vec{r},t)g_{s}(\vec{r},\vec{r}';t-t') J_{s}(\vec{r}',t')\Big] \nonumber\\&& S_{eff.}[e^{a}_{\mu},A^{a,b}_{\mu}] = \int_{-\infty}^{\infty}\,dt' L_{eff.}[e^{a}_{\mu}(\vec{r},t'),A^{a,b}_{\mu}(\vec{r},t')]\nonumber\\&&\approx\int\,dt \int d^3r\sqrt{-det[g_{\mu,\nu}]}\Big[(\frac{\theta}{1536\pi^2})\frac{1}{2}\epsilon^{\mu,\nu,\alpha,\beta}F^{a}_{\alpha,\beta,b}F^{a}_{\mu,\nu,b}   +e_{a, \alpha}e^{\alpha}_{b}F^{a,b}_{\mu,\nu}B(\mu)-\kappa_{eff.}\Big]       \nonumber\\&&
\kappa_{eff.}=\kappa+\mathbf{E}(g,|\varphi|) \approx 0\nonumber\\&&
\end{eqnarray}
The  second term in the first equation  gives  the Green's functions    $ g_{s}(\vec{r},\vec{r}';t-t')$ which is related to the Green's $ \hat{G}_{s}(\vec{r},\vec{r}';t-t';g^{\mu,\nu})$  computed for the    metric $ g^{\mu,\nu}$.
\begin{equation}
 \hat{G}_{s}(\vec{r},\vec{r}';t-t';g^{\mu,\nu})=-i\langle0 |T\Big[ u_{s}(\vec{r},t)u_{s}(\vec{r}',t')\Big] |0\rangle
\label{eqd}
\end{equation}
The Green's function  $ \hat{G}_{s}(\vec{r},\vec{r}';t-t';g^{\mu,\nu})$ obeys a    differential equation which  depends explicitly on the metric tensor. 
\begin{equation}
\frac{1}{\sqrt{|det[g_{\mu,\nu}|}} \partial_{\mu}\Big[\sqrt{|det[g_{\mu,\nu}|}g^{\mu,\nu}\partial_{\nu} \Big] \hat{G}_{s}(\vec{r},\vec{r}';t-t';g^{\mu,\nu})=-\delta[\vec{r}-\vec{r}']\delta[t-t']
\label{diferential}
\end{equation}
When the  the external current  $J_{s}(\vec{r},t)$ vanishes  we can analyze the effective action.  We observe 
that the \textbf{virbein} $\vec{e}$ and the \textbf{field strength} $ F^{a,b}_{\mu,\nu}$ gives rise to  the \textbf{scalar curvature} $R$, $R=e_{a, \alpha}e^{\alpha}_{b}F^{a,b}_{\mu,\nu}$.
We note that the first term $ \sqrt{-det[g_{\mu,\nu}]}(\frac{\theta}{1536\pi^2})\frac{1}{2}\epsilon^{\mu,\nu,\alpha,\beta}F^{a}_{\alpha,\beta,b}F^{a}_{\mu,\nu,b}$ is a total  divergence and therefore as expected  does not  affect the equation of motion $\frac{\delta S_{eff.}}{\delta g_{\mu,\nu}}=0$.  The variation of the second    term $\sqrt{-det[g_{\mu,\nu}]}\Big[e_{a, \alpha}e^{\alpha}_{b}F^{a,b}_{\mu,\nu}B(\mu)-\kappa_{eff.}\Big]$  gives :
\begin{equation}
(R_{\mu,\nu}-\frac{1}{2}Rg_{\mu,\nu})B(\mu)+\kappa_{eff.} g_{\mu,\nu}=0;\hspace{0.1 in}  \kappa_{eff.}\approx 0
\label{eqmotion}
\end{equation}
The term $ G_{\mu,\nu}\equiv  R_{\mu,\nu}-\frac{1}{2}Rg_{\mu,\nu} $  represents the Einstein curvature in elasticity  which coincide with the disclination density  $\Theta_{\mu,\nu}= G_{\mu,\nu}$   (for the spatial coordinates )  \cite{Kleinert}.

\vspace{0.2 in}

\textbf{IV-The  effect of the  elastic  action  given by equation $(18)$   on the elastic field  $\vec{u}(\vec{r},t)$} 

\vspace{0.2 in}

The  Green's function given by  equation $(19) $  must be averaged with respect all the  possible  metrics.  The averaged Green's function  with respect all the possible metrics    The Green's function $g_{s}(\vec{r},\vec{r'};t-t')$ can be obtained from   $\hat{G}_{s}(\vec{r},\vec{r'};t-t';g^{\mu,\nu})$ after performing  the statistical average with respect the metric
$g^{\mu,\nu}$.
\begin{eqnarray}
&&g_{s}(\vec{r},\vec{r'};t-t')=\int  D e^{a}_{\mu} DA^{a,b}_{\mu}  
e^{-i\int_{-\infty}^{\infty}\,dt' L_{eff.}[e^{a}_{\mu}(\vec{r},t'),A^{a,b}_{\mu}(\vec{r},t')]}  \hat{G}_{s}(\vec{r},\vec{r'};t-t';g^{\mu,\nu})\nonumber\\&&
\end{eqnarray}
Therefore one has to obtain the equation of motion for all the possible metrics and to sum the contributions according to their statistical weight  $ e^{-i\int_{-\infty}^{\infty}\,dt' L_{eff.}[e^{a}_{\mu}(\vec{r},t'),A^{a,b}_{\mu}(\vec{r},t')]}$. 

\vspace{0.2 in}

\textbf{V-The metric solution for the elasticity instantons} 

\vspace{0.2 in}

In this section we will compute the metric tensor which emerges from the action   $\int_{-\infty}^{\infty}\,dt' L_{eff.}[e^{a}_{\mu}(\vec{r},t'),A^{a,b}_{\mu}(\vec{r},t')]$. (In the long wave limit  the  term   $\kappa_{eff.} g_{\mu,\nu}$ is replaced  by a  constant term $\kappa_{eff.}\approx 0 $,  which will give rise to a  constant correction to the  free energy of the crystal.) The  effective  action contains two terms: the first term 
  $\sqrt{-det[g_{\mu,\nu}]}(\frac{\theta}{1536\pi^2})\frac{1}{2}\epsilon^{\mu,\nu,\alpha,\beta}F^{a}_{\alpha,\beta,b}F^{a}_{\mu,\nu,b}$ which  is a total  divergence  and   does not  affect the equation of motion; the second term    is given by $\sqrt{-det[g_{\mu,\nu}]}e_{a, \alpha}e^{\alpha}_{b}F^{a,b}_{\mu,\nu}B(\mu)$, the variation of this term generates the    equation of motion   obtained in eq. $(20)$. The saddle point solution    $(R_{\mu,\nu}-\frac{1}{2}Rg_{\mu,\nu})=0$ is  satisfied  for solution which  obey  the self dual or anti-dual conditions for the spin connection and  curvature \cite{Egouchi}:
\begin{equation}
\tilde{A}^{a,b}=\pm A^{a,b} ; \hspace{0.2 in}   \tilde{R}^{a}_{b}=\pm R^{a}_{b}  ;\hspace {0.2 in} \tilde{A}^{a,b}=\frac{1}{2}\sum_{c=0,d=0}^{c=3,d=3}\epsilon_{abcd}A^{c,d}; \hspace{0.2in}
\tilde{R}^{a}_{b}=\frac{1}{2}\sum_{c=0,d=0}^{c=3,d=3}\epsilon_{abcd}R^{c}_{d}
\label{dual}
\end{equation}
Following refs. \cite{Egouchi,Freund} we find  that $\sqrt{-det[g_{\mu,\nu}]}\Big[(\frac{\theta}{1536\pi^2})\frac{1}{2}\epsilon^{\mu,\nu,\alpha,\beta}F^{a}_{\alpha,\beta,b}F^{a}_{\mu,\nu,b}+ e_{a, \alpha}e^{\alpha}_{b}F^{a,b}_{\mu,\nu}B(\mu)\Big]$ admits infinite number of  instantons solutions with different \textbf{ instantons} winding  number.
The duality equation  are  characterized  by an integration factor \textbf{$\Lambda$}. For each $Lambda$  the contribution to the ground state is  given by \textbf{ $e^{\frac{-\Lambda^2}{const}}$}. Therefore, the actual green's function for  lattice dynamics will  involve a summation over  solutions with  different $\Lambda$.  Clearly the dominant solutions occurs for $\Lambda=0$ which corresponds to  an undeformed crystal.

For the remaining part we will  follow refs.  \cite{Freund,Egouchi}  and  solve the duality equations to compute  the metric.
For  an undeformed crystal  the   \textbf{ Lorentz}  transformation  with    the speed of sound  $c=1$ (transversal or longitudinal) is given by:
$ds^2=-c^{2}dt^2+dx^2+dy^2+dz^2$.  Following  ref. \cite{Freund} we introduce  the  four  Euclidean space   coordinates  ($\tau \rightarrow i t$) :  $ds^2=d\tau^2+dx^2+dy^2+dz^2$. Using the  polar coordinates  for a $  S^3$  sphere  of radius   $R^2=\tau^2+x^2+y^2+z^2\equiv \tau^2+r^2 $   
($r$ is the space  radius for  the $S^2$ sphere).  We  parametrize the $  S^3$  sphere   in terms of the space-time   polar coordinates  $\psi$, $\theta$ ,$\phi$ and radius $ R$: 
\begin{eqnarray}
&&x=R\sin[\psi]\sin[\theta]\cos[\phi]; y=R\sin[\psi]sin[\theta]sin[\phi];
z=R\sin[\psi]\sin[\theta]; \tau=R\cos[\psi]\nonumber\\&&
0\le\theta\le\pi;0\le\phi\le2\pi; 0\le\psi\le4\pi
\nonumber\\&&
\end{eqnarray}
The metric tensor for the four dimensional space is  given by   the  orthonormal \textbf{vierbein} basis by  $\Theta^{a}$, $a=0,1,2,3$:
\begin{eqnarray}
&& ds^2=d\tau^2+dx^2+dy^2+dz^2=dR^2+R^2\Big[d\psi^2+sin^2[\psi](   d\theta^2+\sin^2[\theta]d\phi^2)\Big] \equiv  \sum_{a=0}^{3}(\Theta^{a})^2  \nonumber\\&&
\Theta^{0}=dR \nonumber\\&& 
\Theta^{1}= R\Big[\sin[\psi]d\theta-\sin[\theta]\cos[\psi]d\phi\Big]\nonumber\\&&
\Theta^{2}= R\Big[-\cos[\psi]d\theta-\sin[\theta]\sin[\psi]d\phi\Big]\nonumber\\&&
\Theta^{3}= R\Big[d \psi+\cos[\theta]d\phi\Big]\nonumber\\&&
\end{eqnarray}
\textbf{The effect of the crystal deformation $\vec{W}(\vec{r},t)$  replaces  on    the  $S^3$ sphere    the  vierbein     $\Theta^{a}(R,\psi,\theta,\phi)$  by  the deformed crystal  vierbein  $ e^{a}(R,\psi,\theta,\phi)$}
\begin{eqnarray}
&& e^{0}(R,\psi,\theta,\phi)=f(R)\Theta^{0}\nonumber\\&&
e^{1}(R,\psi,\theta,\phi)=g(R)\Theta^{1}\nonumber\\&&
e^{2}(R,\psi,\theta,\phi)=
g(R)\Theta^{2}\nonumber\\&&
e^{3}(R,\psi,\theta,\phi)=
f(R)\Theta^{3}\nonumber\\&&
\end{eqnarray}
The functions $f(R)$ and $g(R)$ are  determined with the help of the duality  relations  for the spin connections  and Ricci curvature  \cite{Egouchi,Freund} .
For the anti-duality condition  we have   : $ \tilde{A}^{a,b}=-A^{a,b}$ , $\tilde{F}^{a}_{b}=- F^{a}_{b}$ ( see  table $1$ of  ref. \cite{Freund}).
The  spin connection $ A^{a,b}$ is obtained  from the exterior derivative $ d(e^{a})$  using the vanishing torsion condition,$ d(e^{a})+ A^{a}_{b} \wedge e^{b}=0$, $ A^{a}_{b}= \delta^{b}_{c}A^{a,c}$ .  The curvature  $R^{a}_{b}$is  obtained from the exterior derivative of the spin connection, $F^{a}_{b}=d(A^{a}_{b})+A^{a}_{c} \wedge A^{c}_{b}$, where $  \wedge$ is the \textbf{wedge} product \cite{Nakahara}.
The anti- duality equations determines  the functions  $f(R)$ and $g(R)$ with an arbitrary integration constant $ \Lambda$.
\begin{eqnarray}
 &&ds^2=f^2(R)\Big [(\Theta^{0})^2+(\Theta^{3})^2\Big]+ \Big[(\Theta^{1})^2+(\Theta^{2})^2\Big] g(R)\nonumber\\&&
 f^2(R)=(1+ \frac{\Lambda R^2}{6})^{-2}\nonumber\\&&
g(R)= f(R)=(1+ \frac{\Lambda R^2}{6})^{-1}\nonumber\\&&
R^2=\tau^2+x^2+y^2+z^2\equiv \tau^2+r^2\nonumber\\&&
\end{eqnarray}  
The solution with  $\Lambda=0$ corresponds to the undeformed crystal.
Next we   change the variables:
\begin{equation}
U=\frac{1}{R}\equiv\frac{1}{\sqrt{r^2+\tau^2}}
\label{uu}
\end{equation}
 The limit  $U\rightarrow 0$ corresponds to the physics at  long distances.
In the new variables  the metric is given by:
\begin{equation}
ds^2=(U^2+\frac{\Lambda}{6})^{-2}\Big[dU^2+\frac{1}{4}U^2(d\psi^2+cos[\theta] d\phi^2)\Big]+
(U^2+\frac{\Lambda}{6})^{-1}\frac{1}{4}\Big[d\theta^2+sin^2[\theta]d\phi^2\Big]
\label{metricg}
\end{equation}
 We observe that the singularity for $U\rightarrow 0$  is removable . For fixed values  $\theta$ and $\phi$ and   $U\rightarrow 0$ the metric becomes  $ds^2\approx (\frac{\Lambda}{6})^{-2}[dU^{2}+\frac{1}{4}U^{2}d\psi^2]$  and   is equivalent to the  two dimensional metric. We perform a change of variable $\hat{V}=U(\frac{\Lambda}{6})^{-1}$ and $\varphi=\frac{\psi}{2}$.  As a result the metric becomes
$ds^2= [d\hat{V}^{2}+\hat{V}^{2}d\varphi^{2}]$.  We observe that  the  singularity $\hat{V}\rightarrow 0$  is removed   if   $\varphi $  is  restricted to   $0<\varphi<2\pi$.  This implies that the value of $\psi=4\pi$ is excluded, $0<\psi<4\pi$.  Following \cite{Kleinert}  we observe that  the   metric  describes a disclination    for  the space time variables  $[U,\psi]$.

The metric  in eq.$(29)$  can be  written in  terms of  the Euclidean  time $\tau$ and  space coordinate $r$ : 
\begin{equation}
\left(\begin{array}{cc}dU\\d\psi\end{array}\right)=
\left[\begin{array}{rrr}
-U^3\tau ,  &   -U^3r     \\
 -U^2r     ,  &  +U^2 \tau\\
\end{array}\right]
\left(\begin{array}{cc} d\tau \\ dr
\end{array}\right)
\end{equation}
In the next step we map the Euclidean metric back  to the Lorentz metric  $\tau\rightarrow it$. 
We replace $U^2=\frac{1}{r^2+\tau^2}\rightarrow
\frac{1}{r^2-t^2}$.
 In order to capture the effect of the disclination we will study     the large scale limit,  $U^2\rightarrow 0 $  and  have two different  regions  $r^2<t^2$  or $r^2>t^2$:
 
\textbf{a) For $t^2>r^2$} we use the   set of variables $V$, $\chi$ :
\begin{equation}
 V= \frac{1}{\sqrt{-r^2+t^2}};\hspace{0.1 in}  r=\frac{1}{V}\sinh[\chi] ; \hspace{0.1 in} t= \frac{1}{V}\cosh[\chi]
\label{cordinate1}
\end{equation}
and  find the metric :
\begin{eqnarray}
&&ds^2=(V^2+\frac{\Lambda}{6})^{-2}\Big[a^2_{V}dV^2- a^2_{\chi}V^2d\chi^2\Big]+(V^2+\frac{\Lambda}{6})^{-1}\frac{1}{4}\Big[d\theta^2+sin^2[\theta]d\phi^2\Big]\nonumber\\&&
 a^2_{V}=\frac{1}{4}e^{-2\chi} ;
a^2_{\chi} =\frac{1}{4}(\frac{5}{2}e^{2\chi}+\frac{3}{2}e^{-2\chi})\nonumber\\&&
\end{eqnarray}
\textbf{b) For $t^2<r^2$} we use a second  set of  variables $Z$,$\eta$ :
\begin{equation}
 Z= \frac{1}{\sqrt{r^2-t^2}};\hspace{0.1 in}  r=\frac{1}{Z}\cosh[\eta] ; \hspace{0.1 in} t= \frac{1}{Z}\sinh[\eta]
\label{cordinate2}
\end{equation}
and  find the metric:
\begin{eqnarray} 
&&ds^2=(Z^2+\frac{\Lambda}{6})^{-2}\Big[-a^2_{Z}dZ^2- a^2_{\eta}Z^2d\eta^2\Big]+(Z^2+\frac{\Lambda}{6})^{-1}\frac{1}{4}\Big[d\theta^2+sin^2[\theta]d\phi^2\Big]\nonumber\\&&
 a^2_{Z}=\frac{5}{8}e^{2\eta}+\frac{3}{8}e^{-2\eta}; 
a^2_{\eta} =\frac{5}{8}e^{-2\eta}+\frac{3}{8}e^{2\eta}\nonumber\\&&
\end{eqnarray}
To conclude this section we observe that the instantons provide a metric  for different values of  the integration constant  $\Lambda$  given in equation  $(27)$ . For each value    $\Lambda$   the  instanton   weight  function  is given by  $ e^{\frac{-\Lambda^2}{const}}$.  For each value of $\Lambda$ we compute  the metric tensor  which affect    the propagation of the  phonons.  To compute  the    propagation  of the phonons  we need to sum up the propagations for different values of $\Lambda$  according to their weight.

\vspace{0.2 in}

\textbf{VI-Bending of Sound  waves as a probe for  Topological Superconductors}

\vspace{0.2in}

In this section we will consider  a simplified solution for the phonon propagation. We will not sum up all the contributions (for different $\Lambda$), instead  we will  consider a fixed value of $\Lambda\neq0$.     We  will 
 derive the  phonon  propagation   at  long distances using  the metric given in eq.$(34)$.
\begin{equation}
ds^2=(Z^2+\frac{\Lambda}{6})^{-2}\Big[-a^2_{Z}dZ^2- a^2_{\eta}Z^2d\eta^2\Big]+(Z^2+\frac{\Lambda}{6})^{-1}\frac{1}{4}\Big[d\theta^2+sin^2[\theta]d\phi^2\Big]\equiv g_{\mu,\nu}dx^{\mu}dx^{\nu}
\label{eq}
\end{equation}
 When  an external stress is  applied  a     phonon   $\vec{u}_{ext.}(\vec{r},t)$  is excited   in  the crystal.  The   action for this excitation    is given by  :
\begin{equation}
L_{waves}=g^{\mu,\nu}\partial_{\mu}u_{ext.}^{a}(\vec{r},t)\partial_{\nu}u_{ext.}^{b}(\vec{r},t)\lambda_{\mu,a,\nu,b}
\label{phonon}
\end{equation}
Here $\lambda_{\mu,a,\nu,b}$ is given by the elastic tensor  given in equation $(5)$.
 We will use the \textbf{Eikonal} approximation and consider \textbf{phonons} instead of sound waves.
In the \textbf{Eikonal} approximation $\partial_{\mu}u_{ext.}^{a}$ is replaced by the four momenta $k^{\mu}$  and  polarization  $k^{\mu}_{a}\rightarrow\frac{d x^{\mu}_{a}}{ds}$.  A further simplification is obtained for  scalar waves.
The Lagrangian  in the \textbf{Eikonal} scalar  approximation  is  replaced  by 
\begin{equation}
L_{Eikonal-scalar}=g_{\mu,\nu}\frac{d x^{\mu}}{ds}\frac{d x^{\nu}}{ds}
\label{scalar}
\end{equation}
The variation of the Lagrangian  $L_{scalar-Eikonal}$ in eq. $(37)$ gives   the equations  of motion :

$\frac{d^2x^{\mu}}{ds^2}+\Gamma^{\mu}_{\nu,\gamma}\frac{dx^{\nu}}{ds}\frac{dx^{\gamma}}{ds}=0$
where
 $\Gamma^{\mu}_{\nu,\gamma}=\frac{1}{2}g^{\mu,\sigma}\Big[\partial_{\gamma}g_{,\nu}+\partial_{\nu}g_{\sigma,\gamma}
-\partial_{\sigma}g_{\nu,\gamma}\Big]$
is the Christofel tensor.
The equations of motion in the large limit  $r^2>t^2$,$ r>>1$  are  given by  equation $(38)$  for the components ,$\eta$, $Z$, $\phi$  and $\theta$
\begin{eqnarray}
&&\frac{d^2x^{\mu}}{ds^2}+\Gamma^{\mu}_{\nu,\gamma}\frac{dx^{\nu}}{ds}\frac{dx^{\gamma}}{ds}=0\nonumber\\&&
\frac{d^2 \eta}{ds^2}+\frac{2}{Z}g(Z)\frac{d\eta}{ds}\frac{d\eta}{ds}=0 ;
\frac{d^2 Z}{ds^2}+Zf(z)\frac{d\eta}{ds}\frac{dZ}{ds}=0 \nonumber\\&&
 g(Z)=1-\frac{2 Z^2}{Z^2+\frac{\Lambda}{6}};
 f(z)=\frac{a^2_{\eta}}{a^2_{Z}} g(Z) \nonumber\\&&
\frac{d^2 \phi}{ds^2}+\frac{2}{\tan(\theta)}(\frac{d\theta}{ds}+\frac{d\phi}{ds})=0;\frac{d^2 \theta}{ds^2}+\frac{sin(2\theta)}{2}\frac{d\phi}{ds}\frac{d\phi}{ds}=0\nonumber\\&&
\end{eqnarray}
 Phonons will travel  along   \textbf{null geodesics} \cite{Visser}.  The  space-time symmetry of the metric  allows to identify   the \textbf{Killing} vectors \cite{Visser}, the directions where the  metric is invariant .
We  find from the metric in  equation $(35)$ that  we have two such Killing vectors, $\phi$  and the temporal coordinate $\eta$.
We consider  the  two Killing vectors and  introduce the integration  of motion $k_{\phi}$ (related to the energy)   and $k_{\eta}$ (the angular momentum  $L_{z}$). 
\begin{equation}
g_{\eta,\eta}\frac{d\eta}{ds}=k_{\eta};\hspace{0.2 in} g_{\phi,\phi}\frac{d\phi}{ds}=k_{\phi}
\label{killing}
\end{equation}
with the metric 
$g_{\eta,\eta}=\frac{-Z^2}{ (Z^2+\frac{\Lambda}{6})^{2}}$; 
$g_{\phi,\phi}=\frac{1}{ 4 (Z^2+\frac{\Lambda}{6}) }sin^2(\theta)$.
 We take the ratio   of the two  functions  in eq.$(39)$   and  find an equation as a function of the time  $\eta$  given by the first expression  in eq.$(40)$. From  eq.$(38)$ we find the second and third  expressions   in eq.$(40)$.  From    the \textbf{null geodesic} condition
$g_{\mu,\nu}\frac{dx^{\mu}}{d\eta}\frac{dx^{\nu}}{d\eta}=0$ we find the last formula  in  eq.$(40)$. 
\begin{eqnarray}
&&\frac{d\phi}{d\eta}=-\frac{4\kappa (Z^2(\eta)+\frac{\Lambda}{6})}{Z^2(\eta)sin^2(\theta)};\kappa\equiv\frac{k_{\phi}}{k_{\eta}}\nonumber\\&&
\frac{d Ln[\frac{d\eta}{ds}]}{d\eta}\frac{dZ(\eta)}{d\eta}+\frac{d^2Z(\eta)}{d\eta^2}+Z(\eta)f(Z[\eta])=0 ; f(Z[\eta])=\frac{a^2_{\eta}}{ a^2_{Z}}g(Z[\eta]) \nonumber\\&&
\frac{d Ln[\frac{d\eta}{ds}]}{d\eta}+2g(Z[\eta])\frac{d Ln[Z(\eta)]}{d\eta}=0; g(Z[\eta])=1-\frac{2Z^2}{Z^2+\frac{\Lambda}{6}} \nonumber\\&&
-\frac{4}{Z^2(\eta)+\frac{\Lambda}{6}}\Big[(\frac{dZ(\eta)}{d\eta})^2+Z^2(\eta)\Big]+\Big[(\frac{d\theta}{d\eta})^2+sin^2(\theta)(\frac{d\phi}{d\eta})^2\Big]=0\nonumber\\&&
\end{eqnarray}
 Using  the first expression  in eq.$(40)$  we obtain  the differential equation for  $ \phi$  given in   eq.  $(41)$. Using  the second and third expression  in eq.$(40)$   allows to    identify  eq. $(41)$ for $Z$.  From the last expression in eq.$(40)$ we identify   eq.$(41)$ for $\theta $.
\begin{eqnarray}
&&\frac{d\phi}{d\eta}=-\frac{4\kappa (Z^2(\eta)+\frac{\Lambda}{6})}{Z^2(\eta)sin^2(\theta)}\nonumber\\&&
\frac{d^2Z }{d\eta^2}    -2g(Z[\eta])\frac{dLn[Z(\eta)]}{d\eta} +   
Z(\eta)f(Z[\eta])=0\nonumber\\&&
\frac{d\theta}{d\eta}-\sqrt{\frac{4}{Z^2(\eta)+\frac{\Lambda}{6}}[(\frac{dZ(\eta)}{d\eta})^2+Z^2(\eta))]-(\frac{(4\kappa (Z^2(\eta)+\frac{\Lambda}{6}))^2}{Z^2(\eta)sin^2(\theta)}}=0\nonumber\\&&
\end{eqnarray}
\textbf{Numerical solution of eq. $(41)$  for short times and large values of $r$.}
We  consider   the initial conditions for  the phonon field at $t=0$.  We assume that an external stress is applied at  a  location $x>1$ ( given in dimensionless units) and $y=z=0$ ,  $[x=r(0)>1,y=0,z=0]$.  The initial propagation velocity  of  the phonon is taken to be  in the $ y$ direction and  is given by  the sound velocity $c=1$ . Using the relations   $dy=xd\phi$,$dy\approx dr=\frac{1}{Z(0)}d\eta$ we find   $\frac{d\phi}{d\eta}|_{\eta=0}=1$  which gives     $\kappa=1$.  In addition we have the initial condition for $Z(0)$ , $Z(0)=\frac{1}{x(0)}<1$ and  the velocity  $\frac{dZ}{d\eta}|_{\eta=0}=-\frac{1}{Z(0)}$.  The initial condition for the  polar angles  are: $\theta(0)=\frac{\pi}{2}$ , $\phi(0)=0$.

The  numerical solution  for short times with  the initial conditions $ x(0)>1$ , initial velocity $ v_{y}(0)=1$ and $\frac{\Lambda}{6} <0.1$:   
 Figures  $1-5$  shows   the  phonon path for the given initial conditions, the full lines describes the bended path and the dashed lines describes the free phonon propagation  ($\Lambda=0$).   We observe that the phonon bends in the  transversal direction to the initial propagation  direction.
Figure $3$ represents the two dimensional $x-y$ view.  The dashed line shows the propagation of the free phonon for a positive velocity in the $y$ direction.  The propagation of the phonon in the curved space  are given by a   phonon full lines . Since the  bending  occurs in  the $x-z$ and $y-z$ space  the  two dimensional x-y view doesn't show the path for the bended  path. 
Figure $4$ shows the  two dimensional   $y-z$ view . Here we see both the free phonon propagation given by  the dashed line  in the $ y$ direction,  the full line shows the phonon propagation in the curved space  which is mostly in the $-z$  direction.   Figure $5$ shows the two dimensional  $x-z$ view. In this view the dashed  free phonon is not seen   ( the $y$  direction is not seen ), we clearly observe the full line which describes the  bended path  for the curved space. 
Figures $1-2$  show the three dimensional view for the bended phonon path ( full line) and free phonon propagation (dashed line). In figure $1$ we observe the bended path in  the three dimensions  for the x-y view.  Similarly fig.$(2)$ shows the free phonon and the propagating phonon in the curved space in three dimensions for the $y-z$ view.

\vspace{0.2 in}

\textbf{Conclusions}

The response  of the  Topological Superconductors  has been studied  with the help of an external stress . Due to the elastic instantons  the metric for the phonon propagation is modified.  When one applies an external stress  the   path   of  the phonon  bends  in the transversal direction to the initial propagation .  Observing the   propagation of the phonon will allow to identify the topological superconductor. 
This results have been obtained using a particular metric with $\Lambda\neq 0$, an exact solution for this problem might involve the  sum over    different values  
for  $\Lambda$. Since the solutions are dominated by the contributions with $\Lambda=0$ a possible way to measure the bending of the phonons in a  Topological superconductor  is to use external stresses to    bend   the crystal  such that the   statistical   weights   with  $\Lambda\neq 0$  will dominate the  partion function.

\clearpage

\begin{figure}
\begin{center}
\includegraphics[width=5.0 in ]{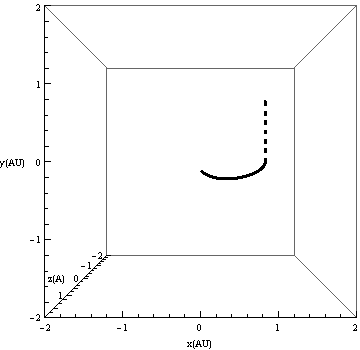}
\end{center}
\caption{x-y view: The dashed line represents the  l propagation of the phonon in the $y$ direction for a flat  metric. The initial conditions are $x(0)>1$ and the velocity is $ v_{y}(0)=1$.  The  full line represents the phonon bending  in the $x<0$  .} 
\end{figure}

\clearpage

\begin{figure}
\begin{center}
\includegraphics[width=5.0 in]{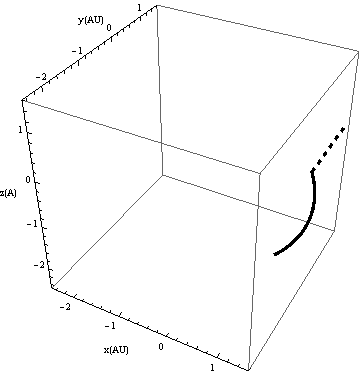}
\end{center}
\caption{y-z view:The dashed line represents the  propagation of the phonon in the $y$ direction for a flat  metric. The initial conditions are $x(0)>1$ and the velocity is $ v_{y}(0)=1$.  The  full line represents the phonon bending  in the $z<0$  direction .} 
\end{figure}

\clearpage

\begin{figure}
\begin{center}
\includegraphics[width=5.0 in]{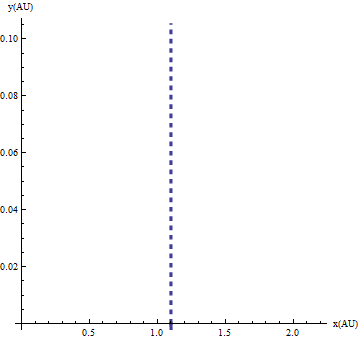}
\end{center}
\caption{Two dimensional y-x view:  Only the  propagation for the flat metric in the $y$ direction is  seen.} 
\end{figure}

\clearpage

\begin{figure}
\begin{center}
\includegraphics[width=5.0 in]{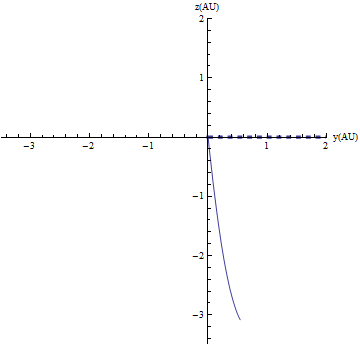}
\end{center}
\caption{Two dimensional y-z view:  The dashed line shows the propagation for a flat metric and the full line shows  the phonon   bending due to the metric} 
\end{figure}

\clearpage

\begin{figure}
\begin{center}
\includegraphics[width=5.0 in ]{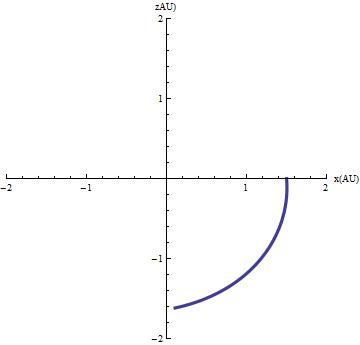}
\end{center}
\caption{Two dimensional x-z view: The propagation  for the flat metric in the $y$ direction is not seen, only the bending of the phonon  due to the metric is shown.} 
\end{figure}

\clearpage



\end{document}